# THE GRAVITATIONAL POTENTIAL OF THE BAR IN NGC 4314

A.C. Quillen,[1,2] Jay. A. Frogel,[1] Rosa A. González[3]

[1] Ohio State University, Department of Astronomy, 174 West 18th Ave., Columbus, OH 43210
[2] E-mail: quillen@payne.mps.ohio-state.edu
[3] Astronomy Department, University of California, Berkeley, CA 94720

## ABSTRACT

We present near-infrared images of NGC 4314 in the $J,H$, and $K$ bands. Colors in these bands are constant across the bar of NGC 4314 indicating that the mass-to-light ratio is constant over much of the galaxy. We have developed an efficient procedure for mapping the gravitational potential of a non-axisymmetric galaxy based on its appearance in the near-infrared using a Fourier Transform method. We account for the thickness of the galaxy's stellar disk by convolving its $K$ image with a function that depends upon the vertical stellar scale height. We apply this procedure to NGC 4314 and find that prograde stellar orbits integrated in the potential are consistent with the shape of the bar for a disk vertical scale height of $h = 350 \pm 100$ pc and a corotation radius of $r = 3.5 \pm 0.5$ kpc.

We find that the inner $m = 4$ Lindblad resonance is approximately at the location of knob-like features seen near the end of the bar. The shape of nearly closed orbits at this radius are lozenge or diamond shaped and have low speeds along the major axis of the bar. The low speeds imply an increase in density along the major axis of the bar. The angle of the major axis of the $m = 4$ component of the surface density begins to twist at the location of its resonance, and the angle of the $m = 2$ component begins to twist at a larger radius near the corotation radius.

*Subject headings:* galaxies: individual (NGC 4314) – galaxies: kinematics and dynamics – galaxies: infrared

## 1. INTRODUCTION

Near-infrared images of galaxies detect light primarily from cool giants and dwarfs that contribute a major fraction of the bolometric luminosity of a galaxy. Particularly in spiral galaxies, these stars are much better traces of the mass distribution of the galaxy than are the bluer, hotter stars (Aaronson 1977; Frogel 1988). Because extinction from dust is far less in the near-infrared than in the optical ($A_K \sim 0.1 A_V$) near-infrared images will also show more accurately the intrinsic shapes of galaxies. For example, near-infrared images can reveal bars that are not easily observed in the optical, e.g. in NGC 1068 (Scoville et al. 1988) and in M82 (Telesco et al. 1991).

Inside the optical disk radius $R_{25}$ (the radius of the 25 mag arcsec$^{-2}$ isophote) the amount of dark matter is only a small fraction of the visible matter, and maximal disk models can account for the observed rotation (Kalnajs 1983, Kent 1986, Begeman 1987). Recent reexamination of the vertical stellar density and velocity distribution of the Milky Way shows that the observations are well fit by a potential due to known matter only (Kuijken & Gilmore 1989a,b,c, Flynn & Fuchs 1994). Therefore, the distribution of near-infrared light will, for many galaxies, be ideally suited for use in dynamical studies to estimate the gravitational potential within $R_{25}$. The ability to obtain near-infrared images with large arrays has led us to formulate a procedure for estimating the gravitational potential of non-axisymmetric galaxies based upon these images.

Previous estimates of gravitational potentials (and rotation curves), based on observed optical images of galaxies with axisymmetric disks have been carried out by Casertano (1983), Kent (1986) and Kalnajs (1983). For barred galaxies, Kent & Glaudell (1989) and Athanassoula et al. (1990) have fit analytical smooth functions to observed optical isophotes in order to study the dynamics of the bar. This procedure has also been used to estimate the form of the $m = 2$ and higher components of the potential in barred galaxies (Athanassoula 1991, Athanassoula & Wozniak 1991). So far as we know no such work has been carried out based upon infrared images.

In §2 we present images in the $J$ (1.25$\mu$m), $H$ (1.65$\mu$m), and $K$ (2.2$\mu$m) bands of NGC 4314, an SBa galaxy with a stellar bar $\sim 130''$ long. The galaxy was classified as peculiar by Sandage (1961) because of a star forming "nuclear ring" at a radius of $\sim 6''$ (Benedict 1992, Benedict et al. 1993, Garcia–Barreto et al. 1991). Our data have improved signal to noise and a larger scale than previously published near-infrared images (Benedict et al. 1992). Comparison of luminosity profiles across the bar in the optical with those in the near-infrared shows that up to 20% of the optical light has been absorbed by the dust lanes along the bar (Benedict et al. 1992). This indicates that optical images of this galaxy do not accurately trace the underlying stellar distribution, and thus are not suitable for estimating the gravitational potential. Uncertainty in the inclination and position angle of the galaxy, and in the form of the bulge will affect the accuracy of an estimate of the gravitational potential. We have chosen NGC 4314 for this study because it is nearly face-on (axis ratio in the outer region $\sim 0.9$) and has a small bulge (extent $< 7''$ Benedict et al. 1992, 1993). In this paper we adopt a distance to NGC 4314 of $D = 10$ Mpc (1 arcsec = 50 pc) (based upon galaxy group membership in the Coma I cloud as discussed in Garcia-Barreto et al. 1991). Our data are a preliminary part of a survey of 200 to 300 galaxies that will produce a library of photometrically calibrated images of late-type galaxies from 0.4 to 2.2$\mu$m.

In §3 we describe the Fourier Tranform method on a grid that we use to estimate the gravitational potential from our data. This procedure is fast and does not rely upon a fit to an assumed form for the bar. Many numerical studies of the dynamics of two and three dimensional model galaxies have previously employed this method to find the potential, (e.g. Efstathiou, Lake & Negroponte 1982). We modify the method by using a convolution function that depends on the thickness of the stellar disk.

In §4, we apply the procedure outlined in §3 to NGC 4314. By comparing the shape of orbits in the potential to the shape of the galaxy, we constrain the vertical scale height of the disk and the bar angular rotation rate. We decompose the potential and the surface brightness of the galaxy into its Fourier components in concentric annuli (e.g. Elmegreen, Elmegreen & Montenegro 1989). We use the axisymmetric component of the potential to predict the rotation curve and the location of resonances. A discussion and summary follow in §5.

## 2. Observations of NGC 4314

The images were taken during photometric conditions on 1993 May 6 and 8 with the 1.8m Perkins Telescope in Flagstaff, AZ using a HgCdTe $256 \times 256$ array in the Ohio State Infra-Red Imaging System (OSIRIS). The array covered a field of $6.6 \times 6.6$ arcminutes, with a spatial scale of 1.50 arcsec/pixel. Individual images were taken with an exposure time of $\sim 2.7$ seconds in $J$ and $K$ and $\sim 1.7$ seconds in $H$. Total on source integration times were approximately 21 minutes in each band. Additionally, the sky was observed for a total integration time that was half of the total on source integration time. Flat fields were constructed from averages of dome flats (in $K$) and from means of skies (in $H$ and $J$). Images were aligned (to the nearest pixel) and combined after a slight non-linearity correction, flat fielding, and sky subtraction. A planar surface was removed from the resulting images to correct for problems in sky subtraction probably due to scattered light.

The final images were calibrated using approximately 12 stars in each of the globular clusters M71 (Frogel, Persson, & Cohen. 1979) and NGC 6712 (Frogel 1985a) to convert to the CTIO/CIT system (Elias et al. 1983). K magnitude Zero points derived from the two globular clusters (which we averaged) agreed to within 0.05 mag. We estimate that our calibration is accurate at this level. In our images the noise per pixel has a variance that corresponds to 21.5 mag per arcsec$^{-2}$ at $K$, 22.0 mag per arcsec$^{-2}$ at $H$, and 23.1 mag per arcsec$^{-2}$ at $J$. The FWHM of stars in the images are $\sim 2''.5$ and are spatially undersampled. Due to the large size of our pixels we cannot study in detail the structure in the region of the nucleus. Images in the three bands and a color map of $J - K$ are displayed in Figure 1. Registration for the $J - K$ color map (Figure 1d) was done by centroiding on stars in the field. Displayed in Figure 3c is a contour map of the outer isophotes of the bar. The end of the bar is at approximately $60''$ (or 3kpc) from the nucleus. We see knobby features near the end of the bar at a radius of approximately $50''$.

### 2.1 Mass-to-Light Ratio & Bulge

We find that the near-infrared colors (see Fig. 1d) are remarkably constant across the bar with $J - K = 0.95 \pm 0.07$ and $H - K = 0.22 \pm 0.07$. This indicates that the stellar population and mass-to-light ratio are constant across the bar. These colors are typical colors of an old stellar population (Frogel et al. 1978), consistent with the findings of Frogel (1985b) and Terndrup et al. (1994) that at infrared wavelengths the bulk of the stellar content of the bulges and inner disks of spiral galaxies are similar to those found in E and S0 galaxies.

The colors variation observed near the nucleus in our near-infrared images can be used to estimate the effect of extinction from dust in the galaxy and the variation in the mass-to-light ratio due to changes in the stellar population. At a radius $r \sim 7''$ at the location of the nuclear ring (Benedict et al. 1993), $J - K$ is higher by $0.15 \pm 0.04$ and $H - K$ is higher by $0.08 \pm 0.04$ than the bar. The colors observed in this ring are consistent with color variations due to extinction assuming the typical galactic extinction ratios $A_K/A_J = 0.38$ and $A_H/A_J = 0.62$ (Mathis 1990). Using these ratios we find that $A_K \sim 0.09$ in the ring, which implies that the $K$ luminosity could be reduced due to extinction by $\sim 10\%$ in this region (assuming a single screen of dust). For dust mixed in with the stars, the $K$ luminosity could be reduced by a much larger factor.

The nucleus itself is also redder than the bar with $J - K$ higher by $0.25 \pm 0.04$ and $H - K$ equal to that of the bar. This may in part be due to a population gradient in the bulge. We see color variations over a small region with size that is consistent with the range of the bulge $r < 7''$ inferred by Benedict et al. (1992) and Benedict et al. (1993) based on the luminosity profile. We note that nowhere in our image (except at radii smaller than the seeing disk) do we see circular isophotes. This is consistent with the small size of the bulge. Using the stellar population models of Worthey (1993) we can limit the change in the mass-to-light ratio in the bulge region (excluding the nuclear ring) to less than 20% assuming that the color change as a function of radius is due entirely to changes stellar population. If there is a decreasing metallicity as a function of radius, then we will overestimate the mass in the nucleus by assuming a constant mass-to-light ratio.

In summary, because of the small size of the bulge relative to our large pixels, we could not reliably subtract it from our images as did Kent & Glaudell (1989 for NGC 936, so we chose not to subtract it. We see color changes only at small radii, and do not correct for changes in the mass-to-light ratio implied by this color change. Our gravitational potential will therefore not be accurate at small radii ($r < 20''$).

## 3. Estimating the Potential

Since we want to use images which are already in a Cartesian grid, the numerical procedure for estimating the gravitational potential that lends itself most easily to the problem is the Fourier Transform Method on a Cartesian grid (Hohl & Hockney 1969; for an introduction see Binney & Tremaine 1987 §2.8). This technique makes use of the observation that the gravitational potential

$$\Phi(\mathbf{x}) = -G \int \frac{\rho(\mathbf{x}')d^3\mathbf{x}'}{|\mathbf{x} - \mathbf{x}'|} \qquad (3.1)$$

can be written as a convolution of the mass density $\rho$ with the function $1/r$. If the galaxy is observed face on, the mass to light ratio is constant across the galaxy, and all the mass is located on a plane, then one may estimate the potential in the plane of the galaxy by convolution from an image of the galaxy. We do the actual computations with a Fast Fourier Transform on a grid with an area four times as large as the galaxy image to remove periodic images and simulate an isolated system (Hohl 1972).

The thickness of the stellar component of galactic disks is, however, not negligible. Optical and near-infrared observations of galaxies that are highly inclined with respect to the line of sight show that the vertical scale height of their disks is constant as a function of radius both in the optical (van der Kruit & Searle 1981a,b, 1982a,b) and in the near-infrared (Wainscoat et al. 1989, Barnaby & Thronson 1992). We can therefore assume that over the spatial range of our images the $z$ dependence of $\rho$ is approximately constant as a function of position in the plane of the galaxy. We write $\rho(\mathbf{x}) = \Sigma(x,y)\rho_z(z)$, where $\Sigma$ is the mass surface density in the plane of the galaxy and $\int_{-\infty}^{\infty} \rho_z(z)dz = 1$. By integrating over $z$ in equation (3.1) we can write the gravitational potential in the plane of the galaxy as

$$\Phi(x,y,z=0) = -G \int \Sigma(x',y') g(x-x', y-y') dx' dy' \qquad (3.2)$$

with the modified convolution function $g(r)$ defined as

$$g(r) = \int_{-\infty}^{\infty} \frac{\rho_z(z)dz}{\sqrt{r^2+z^2}}, \qquad (3.3)$$

where $r$ is the radius in the plane of the galaxy.

If the disk were made up of infinite filaments of constant linear mass density oriented perpendicular to the plane of the galaxy which have potential in the plane $\propto \ln(r)$, then we would convolve the 2 dimensional surface density field with the function $g(r) = \ln(r)$ to estimate the potential in the plane of the galaxy. We expect the galaxy to have a vertical distribution similar to an isothermal disk, with $\rho_z(z) \propto \mathrm{sech}^2(z/h)$ where $h$ is the scale height of the disk. The resulting function $g(r)$ behaves like $\ln(r)$ for $r << h$ and like $1/r$ for large $r$. Figure 2 which plots $g(r)h$ for

$$\rho_z(z) = (1/2h)\mathrm{sech}^2(z/h), \qquad (3.4)$$

shows that $g(r) = 1/r$ for $r >> h$. Since $g(r)$ is flat on scales $r < h$ we do not expect the potential to have structure on these scales. Also shown in Figure 2 is the convolution function for $\rho_z \propto \text{sech}(z/h)$. Since this function is somewhat flatter at small radii than the sech$^2$ convolution function, we expect the resulting potential would also be smoother with less power at scales smaller than $h$. The effect of seeing on an observed image is to convolve the image with a seeing function. This implies that the resulting potential will have also been convolved with the seeing function.

## 4. The Gravitational Potential of NGC 4314

NGC 4314 is close to face–on with an optical axis ratio of $\sim 0.9$ outside the end of bar. Minimal rotational velocities (less than 10 km/sec) were detected in the H$\alpha$ ring (R. Pogge 1994 private communication, Wakamatsu & Nishida 1980). Unfortunately NGC 4314 has no detected HI outside the nucleus (e.g. Huchtmeier 1982), so so we could not use HI measurements either to constrain the inclination of the galaxy or to scale the mass-to-light ratio. As a result we do not correct for the inclination of the galaxy. Because the axis ratio at large radii is close to 1, we do not expect our estimate for the potential to be significantly inaccurate.

For the $z$-distribution of the stars we have assumed a sech-squared law (eq. 3.4) with a vertical scale height $h$ resulting in the convolution function shown in Figure 2 (solid line). We use the $256 \times 256$ $K$ image with the stars removed to estimate the potential. We do the Fast Fourier Transform on a grid four times as large as the image ($1024 \times 1024$). Van der Kruit & Searle (1981a,b) find that the vertical scale height of highly inclined galaxies is related to $H_D$, the exponential disk scale length by $h \sim 1/5 H_D$. Subsequent infrared studies (Barnaby & Thronson 1992, Wainscoat et al. 1989) find that the optical studies overestimated $h$ due to extinction and that $h \sim 1/12 H_D$ for the few galaxies measured. From the optical azimuthally averaged surface brightness of NGC 4314 (Fig. 11 of Benedict et al. 1992) outside the radius of the bar, we estimate an exponential disk scale length of $H_D \approx 90'' = 4.5$ kpc. We began by assuming a vertical scale height consistent with the infrared observations of edge on galaxies, $h = 1/12 H_D = 7''.5 = 375$pc. We integrated prograde stellar orbits with a standard Runge-Kutta integrator for a corotation radius at $r = 70''$ (near the radius of the end of the bar) in the potential resulting from a range of vertical scale heights near this initial estimate. Orbits in the potential with $h > 9''$ were too round to construct the bar from stars in these orbits, while for $h < 5''$ the orbits were more elongated than the bar. We found orbits to be consistent with the appearance of the galaxy for a vertical scale height of $h = 7'' \pm 2'' = 350 \pm 100$pc. The sensitivity of the orbits to the assumed value for the corotation radius is discussed in §4.3.

The potential with vertical scale height $h = 350$pc is shown in Figure 3a which is a contour map of the quantity

$$\Phi(x,y) \left(\frac{M/L_K}{M/L_{K0}}\right)^{-1} \left(\frac{D}{10 \text{Mpc}}\right)^{-1} \qquad (4.1)$$

in units of (km/s)$^2$ or (pc/Myr)$^2$ where $M/L_K$ is the $K$-band mass-to-light ratio with $L_K$ in units of solar $K$ luminosities, and M in solar masses. $M/L_{K0} = 1.236 M_\odot/L_{K\odot}$ is the value predicted by Worthey (1993) for a single burst population model of age 12 Gyr for metalicity [Fe/H]= 0. The noise in our image translates into an uncertainty in the resulting potential of $\pm 21$ (km/s)$^2$. Because of the convolution, the noise in the potential has been smoothed but is also highly correlated. As stated previously (§2.1) our potential is inaccurate near the nucleus ($r \lesssim 20''$) and so we only consider the potential external to this region.

In Figure 3b we have plotted prograde stellar orbits that are nearly periodic in the frame in which the bar is stationary. Points are plotted at equal time intervals in a single orbit. The surface density of the galaxy increases where the velocity of an orbit decreases. For comparison, Figure 3c is a contour map of the K surface brightness drawn to the same scale.

### 4.1 The Components of the Potential

We decompose the potential (and subsequently the surface brightness) into its Fourier components in concentric annuli (e.g. Elmegreen et al. 1989). In Figure 4 we show these components in the plane of the galaxy ($z = 0$) given by

$$\Phi(r,\theta) = \Phi_0(r) + \sum_{m>0} \Phi_{mc}(r)\cos(m\theta) + \Phi_{ms}(r)\sin(m\theta) \qquad (4.2)$$

where $\theta$ is given in the rotating frame in which the bar is fixed and $\theta = 0$ is along the major axis of the bar (at a position angle from north of $145 \pm 5°$). In Figure 4, note the difference in scales along the vertical axis for the $m = 2, 4$ and 6 components. The order of magnitude and form (as a function of $r$) of the $m = 2$ component is similar to those of other bars found by Athanassoula & Wozniak (1991), Athanassoula (1991). Because our convolution function weights the lower frequencies more than the higher frequencies, the resulting potential appears smoother and rounder than the galaxy (compare Fig. 3a with Fig. 3c). The ratio of the $m = 2$ component of the potential to the axisymmetric component is $\sim 0.1$ in the bar whereas this ratio for the galaxy surface brightness is $\sim 1$ (sec §4.4 and Fig. 6). We find that the $m = 2$ and higher order components of the potential are significantly smaller than the axisymmetric component. We note that if the actual scale height of the disk were smaller, the size of the $m = 2$ and higher moments would be larger. We found that a decrease in the scale height of 20% resulted in a $\sim 20\%$ percent increase in the magnitude of the $m = 2$ and 4 components. We do not measure a significant $m = 6$ moment with this data set, but expect that it would be possible to do so with higher signal to noise infrared images. The orbits were integrated in a potential that was the result of a polynomial fit to the moments of the potential. These fits have coefficients listed in Table 1 and are shown in Figure 4.

### 4.2 The Rotation Curve

In order to derive a rotation curve we fit a smooth function to the data points shown in Figure 4 for the axisymmetric component, which we can more easily differentiate than the actual data points themselves. The rotational velocity $v_c = \sqrt{rd\Phi_0/dr}$, and the angular rotation rate ($\Omega = \sqrt{r^{-1}d\Phi_0/dr}$ are shown in Figure 5. Our value for the maximum velocity ($\sim 175$km/s $\times\sqrt{(M/L_K/M/L_{K0})(D/10\text{Mpc})}$ can be compared to that predicted from the Tully Fisher relation. We measure a total magnitude of $7.8\pm0.2$ in the $H$ band from our $H$ image with stars removed. The Tully Fisher relation (following Pierce & Tully (1992) for the $H$ band) predicts a value of $180 \pm 10$ km/s for the circular velocity. Our value for the maximum velocity, $\sim 175$km/s is remarkably consistent with the prediction of the Tully Fisher relation. If there were a component of dark matter in the inner region of the galaxy the circular velocity would be higher than our velocity predicted from the luminosity. A dark matter halo would cause the circular velocity to decrease at large radii more slowly than is seen in Figure 5a.

### 4.3 Resonances

In order to aid in estimating the location of the resonances, we have also plotted in Figure 5b $\Omega \pm \kappa/m$ for $m = 2, 4$, and 6 where the epicyclic frequency $\kappa$ is given by $\kappa^2 = d^2\Phi_0/dr^2 + 3\Omega^2$. By considering the stellar orbits near the corotation radius, Contopoulos et al. (1989) have shown that bars must end interior to this radius, and Kent (1990) finds that barred galaxies have a ratio of corotation radius to bar radius of $\sim 1.2$. For a corotation radius at $r = 70''$ (near the radius of the end of the bar $60''$), we would predict the inner $m = 4$ resonance to be at $r \sim 50''$ and the inner $m = 6$ resonance to be at $r \sim 60''$. We found that the locations of the resonances were insensitive to changes in the vertical scale height. These resonances (Athanassoula 1990) can produce considerable ergodicity (i.e. areas in phase space filled with orbits that are not torri) (Athanassoula 1990). The shapes of the orbits (see Fig. 3b) are consistent with a type I $m = 4$ resonance which has orbits that are rectangular outside the resonance and lozenge (or diamond) shaped inside the resonance (Contopoulos 1988, Athanassoula 1992a) (the sign of the $m = 4$ component of the orbit flips at the resonance). The orbits that have the sharpest peaks occur near the location of $m = 4$ resonance, coincident with the knobby feature near the end of the bar (see Fig. 3c).

An increase in the corotation radius (corresponding to a decrease in the bar angular rotation rate $\Omega_b$) causes the radius of the resonances to increase approximately by the same factor. With a corotation radius of $80''$ ($10''$ larger than assumed previously, see §4.1), we found that the most peaked lozenge shaped orbits are located at a larger radius ($\sim 10''$ larger) than the position of the knobby features. Similarly with a smaller corotation radius of $60''$ we found that the lozenge shaped orbits are within the location of the knobby feature. We therefore find that the morphology of the orbits is consistent with the appearance of the galaxy for a corotation radius of $70'' \pm 10''$. The corotation radius also affects the ellipticity of the orbits. However, our estimate of the vertical scale height (see §4.1) is insensitive to the small uncertainty in the value of the corotation radius.

### 4.4 Comparison to the Moments of the Galaxy

For comparison to Figure 4, plotted in Figure 6 are the components of the surface brightness of the galaxy as observed in $K$ (as defined using equation (4.1) but expanding the surface brightness instead of the potential). We call the moments similarly $S_0$, $S_{mc}$ and $S_{ms}$. Since the components of the potential are the result of convolving the components of the surface brightness with a convolution function, we expect the components of the potential to be smoother than the components of the surface brightness. We note that at the location of the knobby features ($r \approx 50''$) (see Fig. 3c), the azimuthal average $S_0$, has no strong feature, however, the $m = 2, 4$ and $m = 6$ components peak near this radius. The orbits at this radius are diamond (or lozenge) shaped with the sharpest peaks (see Fig. 3b). We suspect that the knobby feature is the result of these diamond shaped orbits near the $m = 4$ resonance.

Because the orbits have low speeds along the bar major axis, there is a density increase along the major axis which causes the $m = 4$ and 6 components of the surface brightness to have the sign that we find. The density increase along the major axis of the bar, and the square shaped orbits outside of the $m = 4$ resonance causes the bar to look "square" (see Fig. 3b,c), with isophotes fit by $|x/a|^c + |y/b|^c = 1$ for $c > 2$ (Athanassoula et al. 1990).

In Figure 7 we have plotted the angle of the major axis of the components of the potential, $\arctan(\Phi_{ms}/\Phi_{mc})$, and of the components of the galaxy surface brightness, $\arctan(S_{ms}/S_{mc})$. A twist in the isophotes caused by the onset of the spiral arms at the end of the bar is seen as a drop in the angle of the major axis of the components. The radius at which the angle of the $m = 2$ component begins to drop is at $60''$–$70''$, for the $m = 4$ component at $30''$–$50''$ and for the $m = 6$ component at $35''$–$50''$. It is interesting that the $m = 2$ component begins to twist at a larger radius than the other components. The onset of the twisting is coincident with the peaks of the $m = 4$ and $m = 6$ components.

In §4.3, we estimated the location of the $m = 4$ resonance to be at a radius of $r \sim 50''$ which is approximately where the the angle of this component of the surface brightness begins to change. In addition the $m = 2$ component begins to twist near the radius of corotation ($\sim 70''$). The $m = 6$ resonance, which occurs outside the $m = 4$ and within corotation, is near where we see a change in the angle of the $m = 6$ component. We find that the angle of the components begins to twist near the location of their resonances. A change in angle in the shape of the orbits (caused by the onset of a density wave) can cause a resonance to be damped. Alternatively, if there is some dissipation in the stellar dynamics, there will also be a twist in angle of the orbit near a resonance. We did not find orbits that when averaged over many rotation periods were offset by a significant angle with respect to the bar. This suggests that the angle changes are due to coherent motion or spiral density waves. It is possible that a $m = 4$ mode spiral density (possibly coupled to the $m = 2$ mode) begins at an smaller radius than the $m = 2$ mode spiral density wave.

## 5. DISCUSSION AND SUMMARY

The constant near-infrared colors observed across NGC 4314 indicate that the mass-to-light ratio is constant across the bar, and that the extent of the bulge is small ($r < 9''$). These results simplify our derivation of the gravitational potential. Using a Fourier transform method, we have estimated the gravitational potential from a $K$ image of the galaxy. We account for the thickness of the stellar disk by convolving the $K$ image with a function that depends upon the vertical stellar scale height. We find that the $m = 2$ and 4 components of the gravitational potential are small compared to the axisymmetric component which suggests that the approximation of a weak bar is a good one except at the location of resonances. We found that prograde stellar orbits integrated in the potential were consistent with the shape of the bar for a vertical scale height of $h = 7'' \pm 2'' = 350 \pm 100$pc and a corotation resonance of $r = 70'' \pm 10''$. This scale height is consistent with the ratio of the vertical scale height to the exponential disk scale length found from near-infrared observations of edge-on galaxies (Barnaby & Thronson 1992, Wainscoat et al. 1989).

We use the axisymmetric component of the potential to predict a rotation curve. We find that the $m = 4$ and $m = 6$ resonances are within the end of the bar and near the location of the knobby features observed. The stellar orbits within the $m = 4$ resonance are lozenge or diamond shaped with low velocities along the major axis of the bar and maximum ellipticity at the location of the knobby features. The low velocities predict a density enhancement along the major axis of the bar which is consistent with the square shape of the isophotes noted by Athanassoula et al. (1990).

We find that the radius at which the position angle of the galaxy surface brightness begins to twist (near the end of the bar) differs in each component. The $m = 4$ component begins to twist at the location of the knobby features near where we predict the $m = 4$ resonance, and the $m = 2$ component begins to twist at a larger radius nearer to the corotation radius. This suggests that the resonances are damped because of coherent angle changes in the orbits. The onset of density waves (spiral arms) could be responsible for damping the resonances. It is possible that a $m = 4$ mode spiral density which may be coupled to the $m = 2$ mode begins at a smaller radius than the $m = 2$ spiral density wave.

In the future, higher quality images could be used to measure the higher order moments of both the potential and the surface brightness of a galaxy. It would then be possible to iterate between predicting the potential from the observed surface density, and constructing the surface density of the bar and spiral arms from orbits which exist in the potential. It would be interesting to compare self-consistent models constructed in this way to the results of N-body simulations, and iteratively constructed theoretical models.

Because there is a large variation in the vertical distribution as a function of stellar age, future work will also consider vertical density distribution functions that differ from the sech$^2$ law used in this paper (see Wainscoat et al. 1989, van der Kruit 1988, Barnaby & Thronson 1992). Functions that are more peaked at $z = 0$ than the sech$^2$ function (such as the exponential function) will result in potentials with more small-scale structure and larger $m = 2$ and higher components. Some barred galaxies are observed to have variations in the vertical distribution such as boxy or peanut shapes when seen nearly edge-on. These shapes, also seen in 3 dimensional numerical simulations of barred galaxies, are due to thickening from vertical resonances and the fire hose instability (Pfenniger & Friedli 1991, Combes et al. 1990, Raha et al. 1991). Our procedure described above does not account for variations in the vertical scale height. However, the vertical oscillation frequency can be estimated using a Fourier Method but with a convolution function that is modified from the one used in this paper. It would be interesting to compare this oscillation frequency to measurements of the stellar velocity dispersion along a bar, such as in Bettonni & Galletta's (1994) study of the barred galaxy NGC 4442. By iteration, it may be also possible to study the vertical structure of the disk.

It is necessary to extend the work of Worthey (1993) which computes mass-to-light ratios for the infrared bands beyond single burst models (e.g. Charlot & Bruzual 1991, 1993). It should become possible to extend the work of Kent (1986) and others which compared optical rotation curves to measured velocities, into the infrared. In particular rotational velocities, predicted from infrared images could be compared to predicted mass-to-light ratios to investigate the quantity of dark matter in the inner region of galaxies.

We hope that the procedure outlined in this paper will be useful as a basis for future dynamical models of gas dynamics in barred and spiral galaxies. In particular, numerical simulations of shocks in barred galaxies (e.g. Athanassoula 1992b) and along spiral arms could be done with more realistic galactic potentials to investigate the response and dissipation of the interstellar medium in these galaxies. The interaction between bars and spiral density waves could also be studied.

We thank the referee, L. Athanassoula, for helpful comments and suggestions which improved the paper. We acknowledge helpful discussions and correspondence with R. Pogge, M. Davies, L. Kuchinski, A. Gould, R. Elston, J. R. Graham, C. Flynn, and D. DePoy. We thank R. Bertram for help with the observations. The OSU galaxy survey is being supported in part by NSF grant AST 92-17716. OSIRIS was built with substantial aid from NSF grants AST 90-16112 and AST 92-18449. J.A.F's research is supported in part by NSF grant AST 92-18281. A.C.Q. acknowledges the support of a Columbus fellowship.

TABLE 1

Polynomial Coefficients[1]

|   | $\Phi_0(r)$ | $\Phi_{2c}(r)$ | $\Phi_{2s}(r)$ | $\Phi_{4c}(r)$ | $\Phi_{4s}(r)$ | $\Phi_{6c}(r)$ | $\Phi_{6s}(r)$ |
|---|---|---|---|---|---|---|---|
| $a_0$ | -9.676e+4 | 5.834e+2 | -4.049e+1 | 1.036e+2 | 2.996e+1 | 1.455e+1 | -1.339e+0 |
| $a_1$ | 2.970e+4 | -4.280e+3 | -8.222e+0 | -1.331e+2 | -1.689e+2 | 1.216e+1 | -2.276e+1 |
| $a_2$ | -4.524e+3 | 1.217e+3 | -8.914e+1 | -2.535e+2 | 1.289e+2 | -1.129e+2 | 2.548e+1 |
| $a_3$ | 3.010e+2 | -1.766e+2 | 3.359e+1 | 1.505e+1 | -4.190e+1 | 1.956e+1 | -1.220e+1 |
| $a_4$ | 1.062e+1 | 5.922e+1 | 6.758e+0 | 4.107e+1 | 1.049e+1 | 9.360e+0 | 4.041e+0 |
| $a_5$ | -1.639e+1 | -1.195e+1 | -3.016e+0 | -9.359e+0 | -1.566e+0 | -2.620e+0 | -6.870e-1 |
| $a_6$ | 4.481e+0 | 5.254e-1 | 1.835e-1 | 3.274e-1 | 5.981e-2 | 1.116e-1 | 3.024e-2 |
| $a_7$ | -4.776e-1 | 6.904e-2 | 1.805e-2 | 6.571e-2 | 8.888e-3 | 1.781e-2 | 3.773e-3 |
| $a_8$ | 1.764e-2 | -5.210e-3 | -1.562e-3 | -4.526e-3 | -6.715e-4 | -1.312e-3 | -3.035e-4 |

[1] Coeficients of the polynomials fit to the functions $\Phi_{mc}(r)$ and $\Phi_{ms}(r)$ (see equation 4.2, Figure 4) used in the integration of orbits. The polynomials are given by $\sum_n a_n r^n$ where r is in kpc with coefficients, $a_n$, in (km/s)$^2$.

# REFERENCES

Aaronson, M. 1977, Ph.D. Thesis, Harvard University

Athanassoula, E. 1990, in "Galactic Models", eds. J. R. Buchler, S. T. Gottesman, & J. H. Hunter, Annals of the New York Academy of Sciences, 596, 181

Athanassoula, E. 1991, "Dynamics of Disc Galaxies", ed. B. Sundelius, Göteburg, Sweden, p. 149

Athanassoula, E. 1992a, MNRAS, 259, 328

Athanassoula, E. 1992b, MNRAS, 259, 345

Athanassoula, E., Morin, S., Wozniak, H., Puy, D., Pierce, M. J., Lombard, J., & Bosma, A. 1990, MNRAS, 245, 130.

Athanassoula, E., & Wozniak, H. 1991 preprint.

Barnaby, D., & Thronson, M. A. 1992, AJ, 103, 41.

Begeman, K. 1987, Ph.D. Thesis, Rijksuniversiteit Groningen

Benedict, G. F., Higdon, J. L., Tollestrup, E. V., Hahn, J. M., & Harvey, P. M. 1992, AJ, 103, 757

Benedict et al. 1993, AJ, 105, 1369

Bettoni, D., & Galletta, G. 1994, A&A, 281, 1

Binney, J., & Tremaine, S. 1987, *Galactic Dynamics* (Princeton U. Press)

Casertano, S. 1983, MNRAS, 203, 735

Charlot, S., & Bruzual A., G. 1991, ApJ, 367, 126

Charlot, S., & Bruzual A., G. 1993, ApJ, 405, 538

Combes, F., Debbasch, F., Friedli, D., & Dfenniger, D. 1990, A&A, 233, 82

Contopoulos, G., Gottesman, S. T., Hunter, Jr., J. H., & England, M.N. 1989, ApJ 343, 608.

Contopoulos, G. 1988, A&A, 201, 44

Elias, J. H., Frogel, J. A., Hyland, A. R., & Jones, T. J. 1983, AJ, 88, 1027

Elmegreen, B., Elmegreen, D. & Montenegro, L. 1989, ApJ, 343, 602

Efstathiou, G., Lake, G., & Negroponte, J. 1982, MNRAS, 199, 1069

Flynn, C., & Fuchs, B. 1994, MNRAS, in press

Frogel, J. A. 1985a, ApJ, 291, 581

Frogel, J. A. 1985b, ApJ, 298, 528

Frogel, J. A. 1988, ARA&A, 26, 51

Frogel, J. A., Persson, S. E., Aaronson, M., & Mathews, K. 1978, ApJ, 220, 75

Frogel, J. A., Persson, S. E., & Cohen, J. G. 1979, ApJ, 227, 449

Garcia-Barreto, J. A., Downes, D., Combes, F., Gerin, M., Magri, C., Carrasco, L., & Cruz-Gonzalez, I. 1991, A&A, 203, 44

Hohl, F. 1972, J. Comput. Phys., 4, 306

Hohl, F., & Hockney, R. W. 1969, J. Comput. Phys., 9, 10

Huchtmeier, W. K. 1980, A&AS, 41, 151

Kalnajs, A. 1983, in Internal Kinematics and Dynamics of Disk Galaxies, IAU Symposium No. 100, ed. E. Athanassoula (Reidel: Dordrecht), p. 87

Kent, S. 1986, AJ, 91, 1301

Kent, S. M. 1990, AJ, 100, 377

Kent, S. M., & Glaudell, G. 1989, AJ, 98, 1588

Kuijken, K., & Gilmore, B. 1989a, MNRAS, 239, 571

Kuijken, K., & Gilmore, B. 1989b, MNRAS, 239, 605
Kuijken, K., & Gilmore, B. 1989c, MNRAS, 239, 651
Mathis, J. S. 1990, ARA&A, 28, 37
Pfenniger, D. & Friedli, D. 1991, A&A, 233, 93
Pierce, M. J., & Tully, R. B. 1992, ApJ, 387, 47
Raha, H., Sellwood, J. A., James, R. A., & Kahn, R. D. 1991, Nature, 352, 411
Sandage, A. 1961, The Hubble Atlas of Galaxies, Carnegie Institution of Washington, Washington D.C
Scoville, N. Z., Matthews, K., Carico, D. P., & Sanders, D. B. 1988, ApJ, 327, L61
Terndrup, D. M., Davies, R. L., Frogel, J. A., DePoy, D. L., & Wells, L.A. 1994, ApJ, 432, 518, in press.
Telesco, C. M., Campins, H., Joy, M., Dietz, K., & Decher, R. 1991, ApJ, 369, 135
van der Kruit, P. C. 1988, A&A, 192, 117
van der Kruit, P. C., & Searle, L. 1981a, A&A, 95, 105
van der Kruit, P. C., & Searle, L. 1981b, A&A, 95, 116
van der Kruit, P. C., & Searle, L. 1982a, A&A, 110, 61
van der Kruit, P. C., & Searle, L. 1982b, A&A, 110, 79
Wainscoat, R. J., Freeman, K. C., & Hyland, A. R. 1989, ApJ, 337, 163
Wakamatsu, K., & Nishida, M. T. 1980, PASJ, 32, 389
Worthey, G. 1993, Ph.D. Thesis, University of California, Santa Cruz

# FIGURE CAPTIONS

**Figure 1.** a) $J$ (1.25 $\mu$m) image in magnitudes per square arcsec. North is up and East is to the left. Offsets are in arcsecs from the position of the nucleus. b) $H$ (1.65 $\mu$m) image in magnitudes per square arcsec. c) $K$ (2.2 $\mu$m) image in magnitudes per square arcsec. d) $J - K$ color image in magnitudes. Black represents redenning. Note how constant the colors are across the bar.

**Figure 2.** The convolution function $g(r)$ for a disk with a sech$^2$ vertical density distribution (equation 3.4) (shown as a solid line), and for a disk with a sech vertical density distribution (shown as dotted line). Plotted is $g(r/h)h$ so that the axes are unitless. For $r >> h$ the convolution functions are equal to $1/r$.

**Figure 3.** a) The potential derived with vertical scale length $h = 7'' = 350$pc. Contours are $10^4$ (km/s)$^2$ apart with the central contour at $-9 \times 10^4$ (km/s)$^2$. Units of the potential are multiplied by factors given in the text. The potential is much smoother and rounder than the galaxy. b) Prograde stellar orbits for a corotation radius at $r = 70''$ in the frame in which the bar is stationary. Points are plotted at equal time intervals in a single orbit. The density of the galaxy should increase where the velocity decreases. c) Contour plot of $K$ surface brightness. The lowest contour is at a level of 19.45 magnitudes per arcsec$^2$. The difference between contours is equal to the surface brightness which corresponds to the magnitude of the lowest contour.

**Figure 4.** The components of the potential shown in Figure 3. $\Phi_0$ is the azimuthally averaged component in (pc/Myr)$^2$ or (km/s)$^2$ times factors given in the text. The $m = 2, 4$ and $6$ components are given divided by $\Phi_0$. For these components the solid triangles show show the cosine components and open squares show the sine components. Polynomial fits to these functions (see Table 1 for coefficients) are shown as solid lines. Note the difference in the vertical scales for the $m = 2, 4$ and 6 components.

**Figure 5.** a) Circular rotation speed. b) Angular rotation rate $\Omega$ is plotted as the solid line. As an aid to finding the resonances, we have also plotted $\Omega \pm \kappa/2$, the dotted lines, $\Omega \pm \kappa/4$, the short dashed lines, and $\Omega \pm \kappa/6$, the long dashed lines. For a bar angular rotation rate $\Omega_b = 0.046$Myr$^{-1}$ (shown as the horizontal line), the corotation radius is at $r = 70''$, the inner $m = 4$ resonance is at $r \sim 50''$ and the inner $m = 6$ resonance is at $r \sim 57''$.

**Figure 6.** The components of the $K$ surface brightness of the galaxy. $S_0$ is the azimuthally averaged component in mag/arcsec$^2$. The $m = 2, 4$ and 6 components are given divided by $S_0$. For these components the solid triangles show show the cosine components and open squares show the sine components.

**Figure 7.** a) Angles (in degrees) of the major axis of the components of the galaxy surface brightness as a function of radius. The solid triangles denote the angle of the $m = 2$ component, the open squares the angle of the $m = 4$ component and the stellar pentagons show the angle of the $m = 6$ component. b) Same as Figure 7a) but with a larger scale. Note that the angle of the $m = 4$ component begins to decreases at an shorter radius than the $m = 2$ component. c) Angles (in degrees) of the components of the potential as a function of radius. Symbols are the same as in Figure 7a).